\documentclass[twocolumn]{revtex4}

\usepackage{titlesec}
\titleformat{\subsection}[runin]{\normalfont\bfseries}{\thesubsection.}{.5em}{}[.]\titlespacing{\subsection}{0pt}{2ex plus .1ex minus .2ex}{.8em}
\titleformat{\subsubsection}[runin]{\normalfont\itshape}{\thesubsubsection.}{.3em}{}[.]\titlespacing{\subsubsection}{0pt}{1ex plus .1ex minus .2ex}{.5em}
\titleformat{\paragraph}[runin]{\normalfont\itshape}{\theparagraph.}{.3em}{}[.]\titlespacing{\paragraph}{0pt}{1ex plus .1ex minus .2ex}{.5em}

\usepackage{amsmath} 
\usepackage{amssymb}
\usepackage{amsfonts}
\usepackage{latexsym}
\usepackage{amsthm}
\usepackage{amsxtra}
\usepackage{amscd}
\usepackage{bbm}
\usepackage{mathrsfs}
\usepackage{bm}
\usepackage{graphicx}

\usepackage{graphicx, color}

\definecolor{darkred}{rgb}{0.9,0,0.3}
\definecolor{darkblue}{rgb}{0,0.3,0.9}
\definecolor{darkgreen}{rgb}{0,0.5,0.0}

\usepackage[pdftex, colorlinks, linkcolor=blue,citecolor=red]{hyperref}

\theoremstyle{plain} 

\newtheorem*{theorem*}{Theorem}

\newtheorem*{lemma*}{Lemma}

\newtheorem*{corollary*}{Corollary}

\newtheorem*{proposition*}{Proposition}

\newtheorem*{definition*}{Definition}

\newtheorem*{conjecture*}{Conjecture}

\theoremstyle{definition} 

\newtheorem*{example*}{Example}

\newtheorem*{remark*}{Remark}

\renewcommand{\b}[1]{\boldsymbol{\mathrm{#1}}} 
\renewcommand{\cal}{\mathcal} 
 
\newcommand{\fra}{\mathfrak} 
\newcommand{\ovl}[1]{\overline{#1} \!\,} 

\newcommand{\E}{\mathbb{E}}

\newcommand{\C}{\mathbb{C}}

\newcommand{\ii}{\mathrm{i}}
\newcommand{\dd}{\mathrm{d}}

\newcommand*{\deq}{\mathrel{\vcenter{\baselineskip0.65ex \lineskiplimit0pt \hbox{.}\hbox{.}}}=}

\renewcommand{\geq}{\geqslant}
\renewcommand{\epsilon}{\varepsilon}

\newcommand{\qq}[1]{[\![{#1}]\!]}

\newcommand{\pb}[1]{\bigl({#1}\bigr)}
\newcommand{\pB}[1]{\Bigl({#1}\Bigr)}

\newcommand{\pBB}[1]{\Biggl({#1}\Biggr)}

\newcommand{\qb}[1]{\bigl[{#1}\bigr]}
\newcommand{\qB}[1]{\Bigl[{#1}\Bigr]}

\newcommand{\qBB}[1]{\Biggl[{#1}\Biggr]}

\newcommand{\abs}[1]{\lvert #1 \rvert}
\newcommand{\absb}[1]{\bigl\lvert #1 \bigr\rvert}
\newcommand{\absB}[1]{\Bigl\lvert #1 \Bigr\rvert}

\newcommand{\avg}[1]{\langle #1 \rangle}

\newcommand{\scalar}[2]{\langle{#1} \mspace{2mu}, {#2}\rangle}

\newcommand{\scalarB}[2]{\Bigl\langle{#1} \,\mspace{2mu},\, {#2}\Bigr\rangle}

\newcommand{\st}[1]{\tilde{ {#1}}}

\DeclareMathOperator{\tr}{Tr}

\DeclareMathOperator{\re}{Re}
\DeclareMathOperator{\im}{Im}

\let\e=\varepsilon

\renewcommand \S {\textbf{S}}
\renewcommand \C {\textbf{C}}
\renewcommand \u {\textbf{u}}
\newcommand \St {\st{\textbf{S}}}
\newcommand \lambdat {\st{\lambda}}
\newcommand \Wishart {\b{\cal W}}

\newcommand{\stj}{\fra g}

\newcommand{\rtr}{\cal R}

\newcommand{\str}{\cal S}

\begin{document}

\title{On the overlaps between eigenvectors of correlated random matrices}
\author{Jo{\"e}l Bun$^{1}$, Jean-Philippe Bouchaud$^2$, Marc Potters$^2$}
\affiliation{$^1$ LPTMS, CNRS, Univ.\ Paris-Sud, Universit{\'e} Paris-Saclay, 91405 Orsay, France}
\affiliation{$^2$ Capital Fund Management, 23--25, rue de l'Universit\'e, 75\,007 Paris}
\begin{abstract}
We obtain general, exact formulas for the overlaps between the eigenvectors of large correlated random matrices, with additive or multiplicative
noise. These results have potential applications in many different contexts, from quantum thermalisation to high dimensional statistics. We find that
the overlaps only depend on measurable quantities, and do {\it not} require the knowledge of the underlying ``true'' (noiseless) matrices.
We apply our results to the case of empirical correlation matrices, that allow us to estimate reliably the width of the spectrum 
of the true correlation matrix, even when the latter is very close to the identity. We illustrate our results on the example of stock returns correlations, that clearly reveal 
a non trivial structure for the bulk eigenvalues. We also apply our results to the problem of matrix denoising in high dimension.  
\end{abstract}
\sloppy
\maketitle

\section{Introduction}

The structure of the eigenvalues and eigenvectors of large random matrices is of primary importance in many different contexts, from quantum 
mechanics to high dimensional data analysis. Correspondingly, Random Matrix Theory (RMT) has established itself as a major discipline, at the frontier 
between theoretical physics, mathematics, probability theory, and applied statistics, with a somewhat intimidating corpus of knowledge \cite{akemann2011oxford}. 
One of the most striking applications of RMT concerns quantum chaos and quantum transport \cite{beenakker1997random}, with renewed interest coming from problems of quantum ergodicity 
(``eigenstate thermalisation'') \cite{deutsch1991quantum,ithier2015thermalisation}, entanglement and dissipation (for recent reviews see \cite{nandkishore2015many,eisert2015quantum}). 
In the context of signal processing, RMT is of primary importance in the analysis of high dimensional statistics \cite{johnstone2006high,bouchaud2009financial, paul2014random}, 
wireless communication channels \cite{tulino2004random,couillet2011random}, etc. Other examples cover the dynamics of complex systems -- from random ecologies \cite{may1972will} to glasses and spin-glasses \cite{fyodorov2004complexity}. 

Whereas the spectral properties of random matrices have been investigated at length, the interest has recently shifted to the statistical properties of their eigenvectors -- see e.g.\ \cite{deutsch1991quantum,wilkinson1995brownian}
and \cite{ledoit2011eigenvectors, allez2012eigenvector, bourgade2015fixed, allez2014eigenvector, bloemendal2015principal, couillet2015kernel,monasson2015estimating} for more recent papers. 
In particular, one can ask how the eigenvectors of a sample matrix $\b S$ resemble those
of the \emph{population} (or \emph{pure}) matrix $\b C$ itself. We recently obtained in \cite{bun2015rotational} explicit formulas for the overlaps between these pure and noisy eigenvectors for a wide class of 
random matrices, generalizing results obtained for sample covariance/correlation matrices of \cite{ledoit2011eigenvectors} -- obtained as $\b S = \sqrt{\b C} \b{\cal W} \sqrt{\b C}$ where $\b{\cal W}$ is a Wishart matrix -- 
and for matrices of the form $\b S = \b C + \b W$, where $\b W$ is a symmetric Gaussian random matrix \cite{biane2003free, allez2014eigenvector, allez2014eigenvectors}. In the present paper, we want to generalize these results to the overlaps between the eigenvectors of 
{\it two} different realizations of such random matrices, that remain correlated 
through their common part $\b C$. For example, imagine one measures the sample correlation matrix of the same process, but on two non-overlapping time intervals, 
characterized by two independent realizations of the Wishart noises $\b{\cal W}$ and $\st{\Wishart}$. How close are the corresponding eigenvectors expected to be? We provide exact, 
explicit formulas for these overlaps in the high dimensional regime. Precisely, we give a transparent interpretation to our formulas and generalize them to various cases, in particular when the noises are correlated. Perhaps surprisingly, these overlaps
can be evaluated from the empirical spectrum of $\b S$ only, i.e.\ \emph{without any prior knowledge} of the pure matrix $\b C$ itself. This leads us to propose 
a statistical test based on these overlaps, that allows one to determine whether two realizations of the random matrix $\b S$ and $\st{\S}$ indeed correspond to 
the very same underlying ``true'' matrix $\b C$ either in the multiplicative and additive cases defined above. 
We shall also revisit the theory of rotational invariant estimators (RIE) \cite{james1961estimation} that encountered a lot of attentions recently (see \cite{bun2017cleaning,ledoit2011eigenvectors,bartz2016cross,karoui2008spectrum} to cite a few). 

\section{Theoretical results}

\subsection{Inversion formula}

Throughout the following, we consider $N \times N$ symmetric random matrices and denote by $\lambda_1 \geq \lambda_2 \geq \dots \geq \lambda_N$ the eigenvalues of $\b S$ and by $\b u_1, \b u_2, \dots, \b u_N$ the corresponding eigenvectors. Similarly, we denote by $ \st{\lambda}_1 \geq  \st{\lambda}_2 \geq \dots \geq  \st{\lambda}_N$ the eigenvalues of $\st{\S}$ and by $ \st{\u}_1, \st{\u}_2, \dots, \st{\u}_N$ the associated eigenvectors. Note that we will sometimes index the eigenvectors by their corresponding eigenvalues for convenience. We emphasize that we allow the parameters that describe the randomness of $\S$ and $\st \S$ to be different. The central object that we focus on in this study are the asymptotic ($N \to \infty$) \emph{scaled, mean squared overlaps} 
\begin{equation}
	\label{eq:mso}
	\Phi(\lambda,  \st\lambda) \deq N \E\qb{\scalar{\u_{\lambda}}{{\st{\u}}_{\st\lambda}}^2},
\end{equation}
that remain $\cal O(1)$ in the limit $N \to \infty$. In the above equation, the expectation $\E$ can be interpreted either as an average over different realizations of the randomness or, for fixed randomness, as
an average over small ``slices'' of eigenvalues of width $\eta = {\rm d\lambda} \gg N^{-1}$, such that the result becomes self-averaging in the large $N$ limit. 
We will study the asymptotic behavior of \eqref{eq:mso} using the complex function 
\begin{equation}
	\label{eq:psi_transform}
	\psi_N(z,  \st z) \;\deq\; \E\qBB{\frac1N \tr \left[ (z-\b S)^{-1} ( \st z - \st{\b S})^{-1} \right]},
\end{equation}
where $z,  \st z \in \mathbb{C}$. For large random matrices, we expect the eigenvalues $[\lambda_i]_{i\in\qq{1,N}}$ and $[\st\lambda_i]_{i\in\qq{1,N}}$ to stick to their \emph{classical} locations, i.e. smoothly allocated with respect to the quantile of the spectral density. Differently said, the sample eigenvalues become deterministic in the large $N$ limit. Hence, we obtain after taking the continuous limit that $\psi_N(z,  \st z) \sim \psi(z,  \st z)$ where the limiting value is given by
\begin{equation}
\label{eq:zeta_to_overlap}
\psi(z,\st z) \;\deq\; \int \! \! \! \int \frac{\varrho(\lambda)}{z-\lambda} \frac{\st\varrho(\st\lambda)}{\st z -\st \lambda} \Phi(\lambda, \st\lambda) \dd\lambda \dd\st\lambda,
\end{equation}
with $\varrho$ and $\st\varrho$ the spectral density of $\b S$ and $\st{\b S}$. Then, it suffices to compute
\begin{align}
\label{eq:m value 1}
& \psi(x - \ii\eta, y \pm \ii\eta) \; = \;  \nonumber \\
&\int \! \! \! \int \frac{(x - \lambda + \ii\eta)}{(x-\lambda)^2 + \eta^2}\frac{(y-\st \lambda \mp \ii\eta)}{(y -\st\lambda)^2 + \eta^2} \varrho(\lambda)\st\varrho(\st\lambda)\Phi(\lambda, \st\lambda) \dd\lambda \dd\st\lambda\,, \nonumber
\end{align}
to deduce that
\begin{align}
	& \re\qb{\psi(x - \ii\eta, y + \ii\eta) - \psi(x - \ii\eta, y - \ii\eta) } \; = \; \nonumber \\ 
	& \quad 2 \int\! \! \!\int \frac{\eta \varrho(\lambda)}{(x-\lambda)^2 + \eta^2}\frac{\eta \st\varrho(\st\lambda)}{(y -\st\lambda)^2 + \eta^2}  \Phi(\lambda, \st\lambda) \dd\lambda \dd\st\lambda.
\end{align}
We may now invoke Sokhotski-Plemelj identity to obtain the \emph{inversion} formula
\begin{equation}
\label{eq:inversion_formula}
\Phi(\lambda, \st\lambda) = \frac{\re[\psi_0(z, \ovl{\st z}) - \psi_0(z, { \st z})]}{2 \pi^2 \varrho(\lambda)\st\varrho( \st\lambda)},
\end{equation}
with $z = \lambda - \ii\eta$, $\ovl{z}$ its complex conjugate and $\psi_0 \equiv \lim_{\eta\downarrow0^+}\psi$. This last formula tells us that in the high-dimensional regime, we can study the mean squared overlap \eqref{eq:mso} through the bi-variate function $\psi(z,\tilde z)$ which is easier to handle using tools from RMT (see below). 

\subsection{Multiplicative noise}
\label{sec:multiplicative_noise}

The study of the asymptotic behavior of the function $\psi$ requires to control the resolvent of $\b S$ and $\st{\b S}$ entry-wise. It was shown recently that one can approximate these (random) resolvents entry-wise by deterministic equivalent quantities \cite{burda2004signal,bun2015rotational, knowles2014anisotropic}. We begin first with the Wishart multiplicative noise $\Wishart$ that we introduced in the introduction. More precisely, let $\S := \sqrt{\C} \Wishart \sqrt{\C}$ and $\st\S := \sqrt{\C} \st{\Wishart} \sqrt{\C}$ where $\Wishart$ and $\st{\Wishart}$ are two independent Wishart matrices with possibly two different observation ratios $q \deq N/T$ and $\st{q} = N/\st T$. By independence, we have 
\begin{equation}
	\label{eq:psi_resolvent_decomp}
	\psi_N(z,\st z) = \frac1N \sum_{k,l}^{N} \E_{\cal P}\qb{(z - \b S)^{-1}_{kl}} \E_{\st{\cal P}}\qb{(\st z - \st{\b S})^{-1}_{kl}},
\end{equation}
where $\E_{\cal P}[\cdot]$ (resp. $\E_{\st{\cal P}}[\cdot]$) denotes the expectation value over the probability measure $\cal P$ (resp. $\st{\cal P}$) associated to $\b S$ (resp. $\st{\b S}$). 
Then, we use the \emph{deterministic} estimate of the resolvent of $\S$ which yields for $N \to \infty$ \cite{burda2004signal, bun2015rotational, knowles2014anisotropic}:
\begin{equation}
	\label{eq:AILL_SCM}
\E_{\cal P}\qb{(z - \b S)^{-1}_{kl}} \sim \zeta(z) \pB{z\zeta(z) - \b C}_{kl}^{-1} + O(N^{-1/2}), 
\end{equation}
where we defined 
\begin{equation}
	\label{eq:zeta_MP}
	\zeta(z) \;\deq\; \frac{1}{1-q+qz \stj(z)}\,,
\end{equation} 
with $\stj(z)$ is the Stieltjes transform of $\S$ defined as the limiting value of $N^{-1} \tr[(z - \b S)^{-1}]$.
The estimate \eqref{eq:AILL_SCM} holds as well for $\st{\b S}$ by replacing $\zeta$, $q$ and $\stj$ with $\st\zeta$, $\st q$ and $\st \stj$. Note that we can deduce the fixed-point equation associated to $\stj(z)$ by taking the normalized trace in \eqref{eq:AILL_SCM}, this yields for $N \to \infty$
\begin{equation}
	\label{eq:MP_eq}
	\stj(z) \sim \zeta(z)\stj_{\C}(z\zeta(z))\,,
\end{equation}
where $\stj_{\C}$ is the Stieltjes transform associated to the pure matrix $\C$. Again, the value of $\st\stj$ is obtained from \eqref{eq:MP_eq} by replacing $\zeta$ with $\st\zeta$.

By plugging Eq.\ \eqref{eq:zeta_MP} into Eq.\ \eqref{eq:psi_resolvent_decomp}, we get
\begin{equation*}
\psi_N(z,\st z) \sim \frac1N \tr [ \zeta(z) (z\zeta(z) - \b C)^{-1} \st \zeta(\st z)( \st z\st \zeta(\st z) - \b C)^{-1} ]\,,
\end{equation*}
and then, using the identity 
\begin{align}
	\label{eq:resolvent_identity}
	& (z\zeta(z) - \b C)^{-1} (\st z \st \zeta(\st z) - \b C)^{-1} \; = \; \nonumber \\
	& \quad\frac{1}{\st z \st \zeta(\st z) - z \zeta(z)} [ (z \zeta(z) - \b C)^{-1} - (\st z \st \zeta(\st z) - \b C)^{-1} ]\,,
\end{align}
we obtain
\begin{equation*}
\psi_N(z,\st z) \sim   \frac{\zeta(z)\, \st \zeta(\st z)}{\st z \st \zeta(\st z) - z \zeta(z)} \frac1N \tr [ (z \zeta(z) - \b C)^{-1} -  (\st z \st \zeta(\st z) - \b C)^{-1} ].
\end{equation*}
From this last equation, we deduce 
\begin{align}
\label{eq:psi_SCM_tmp}
& \psi_N(z,\st z) \sim \frac{1}{\st z \st \zeta(\st z) - z \zeta(z)} \times \nonumber \\
& \pBB{ \frac{\st \zeta(\st z)}{N} \tr \qB{ \zeta(z) (z\zeta(z) - \b C)^{-1} }   -  \frac{\zeta(z)}{N}  \tr \qB{ \st \zeta ( \st z) (\st \zeta(\st z) - \b C)^{-1} } }\,. \nonumber
\end{align}
One notices from \eqref{eq:MP_eq} that the two normalized trace terms in the latter equation are exactly given by $\stj(z)$ and $\st\stj(\st z)$ in the large $N$ limit. 
We therefore conclude that in the case of a multiplicative Wishart perturbation, the asymptotic value of \eqref{eq:psi_transform} reads
\begin{equation}
\label{eq:psi_SCM}
\psi(z,\st z) \sim   \frac{\st \zeta(\st z) \stj(z) - \zeta(z) \st \stj(\st z) }{\st z \st \zeta(\st z) - z \zeta(z)}\,,
\end{equation}
which holds for any $q = \cal O(1)$ and $\st q = \cal O(1)$. The striking observation in Eq.\ \eqref{eq:psi_SCM} is that the result does not depend \emph{explicitly} on the population matrix $\C$ that we wish to estimate. This feature is crucial since it indicates that we shall be able to characterize the mean squared overlap \eqref{eq:mso} in terms of observable variables only. 

Now that we have determined the asymptotic value $\psi(z, \st z)$, let us now compute the main quantity of interest, i.e.\ Eq.\ \eqref{eq:mso}. To that end, it is convenient to work with the complex function $m(z) := 1/(z \zeta(z))$. Indeed, by expressing \eqref{eq:psi_SCM} in terms of the function $m$, we end up with
\begin{align*}
	&\psi(z, \st z) \; = \;  \\ 
	& \quad \frac{1}{q \st q z \st z} \left[ \frac{(\st q z - q \st z) \st m^2}{m - \st m} + \frac{(q - \st q)\st m}{m - \st m} \right] + \frac{m+\st m}{q\st z} - \frac{1-q}{qz\st z}\,. \nonumber
\end{align*}
Defining $m_0(\lambda) = \lim_{\eta\downarrow0} m(\lambda - \ii\eta) \equiv m_R(\lambda) + \ii m_I(\lambda)$, one obtains after some elementary computations (see Appendix \ref{app:psi_SCM} for details) the following general result:
\begin{align}
	\label{eq:mso_SCM_general}
	& \Phi_{q,\st q}(\lambda, \st\lambda) = \frac{2(\st q \lambda - q \st\lambda) \alpha(\lambda, \st\lambda) + (\st q - q)\beta(\lambda, \st\lambda)}{\lambda\,\st\lambda \, \gamma(\lambda, \st\lambda) }\,,
\end{align}
where we defined 
\begin{eqnarray}
	\label{eq:mso_SCM_general_aux_var}
	\alpha(\lambda,\st\lambda) & \deq & m_R(\lambda) \abs{\st m_0(\st\lambda)}^2 - \st m_R(\st\lambda) \abs{m_0(\lambda)}^2 \\
	\beta(\lambda,\st\lambda) & \deq & \abs{\st m_0(\st\lambda)}^2 - \abs{m_0(\lambda)}^2  \nonumber\\
	\gamma(\lambda, \st\lambda') & \deq & \qb{(m_R(\lambda) -  m_R( \st\lambda))^2 + (m_I(\lambda) +  \st m_I( \st\lambda))^2 } \times  \nonumber \\
	                          &  & \qb{(m_R(\lambda) -  \st m_R( \st\lambda))^2 + (m_I(\lambda) -  \st m_I( \st\lambda))^2 }\,. \nonumber 
\end{eqnarray}
The final result \eqref{eq:mso_SCM_general} is invariant under the exchange of $(q, \lambda)$ with $(\tilde q, \tilde \lambda)$ and does not depend on $\C$ explicitly, as expected. We will see in the next section that it is a crucial feature in order to establish an observable stability test. 

It is easy to show that this formula reproduces the mean squared overlap between a given sample eigenvector and its true value in the limit $\st q \to 0$ (see e.g.\ \cite{ledoit2011eigenvectors, bun2015rotational}). To prove our claim, let us consider $\st q \to 0$, for which $\st\lambda \to \mu$ where $\mu$ denotes the corresponding population eigenvalue \cite{anderson1963asymptotic}. In that specific framework, we have $\st m_R = 1/\mu$  and $\st m_I = 0$. Hence, we deduce from \eqref{eq:mso_SCM_general} that
\begin{eqnarray}
	\Phi_{q,\st q \to 0}(\lambda, \mu) 
	& = & \frac{q\mu}{\lambda \abs{1-\mu m_0(\lambda)}^2 },
\end{eqnarray}
which is exactly the result of \cite{ledoit2011eigenvectors}. 

Next, we look at the case where we split our datasets in two windows of same size ($q = \st q$) which is relevant when ones wishes to measure the stability of the eigenvectors associated to the same eigenvalue. For $q = \st q$, the eigenvalues $\lambda$ and $\tilde\lambda$ are now distributed according to the same density function so that $\st m(\st\lambda) = m(\st\lambda)$. Moreover, we infer from \eqref{eq:mso_SCM_general_aux_var} that the contribution of $\beta(\lambda, \st\lambda)$ in \eqref{eq:mso_SCM_general} vanishes. The self-overlap limit $\tilde\lambda \to \lambda$ needs to be handled with care as the formula \eqref{eq:mso_SCM_general} seems ill-defined when $q = \st q$. Nevertheless, if we write $\st\lambda = \lambda + \e$ with $\e > 0$, one has 
\begin{align*}
	& \alpha(\lambda, \lambda+\e) \; = \; \e^2 \pb{\abs{m_0}^2 \partial_\lambda m_R  - m_R \partial_\lambda \abs{m_0}^2} + \cal O(\e^3)\,, \nonumber \\
	& \gamma(\lambda, \lambda+\e) \; = \; 4 \e^2 m_I^2(\lambda) |\partial_\lambda m_0(\lambda)|^2 + \cal O(\e^3).
\end{align*}
As a consequence, we conclude by plugging these two expressions into \eqref{eq:mso_SCM_general} and then setting $\e = 0$ that the \emph{self-overlap} is given by
\begin{equation}
	\label{eq:mso_SCM_bulk_same_q_eig}
	\Phi(\lambda,\lambda) = \frac{q}{2 \lambda^2}  
	\frac{ \abs{m_0(\lambda)}^4 \partial_\lambda\qb{m_R(\lambda)/\abs{m_0(\lambda)}^2}}{m_I^2(\lambda) |\partial_\lambda m_0(\lambda)|^2}\,.
\end{equation}

We now explain how we can extend these results to more general multiplicative noise. More specifically, let us consider matrices of the form $\b S = \sqrt{\b C} \b O \b B \b O^* \sqrt{\b C}$, where $\b O$ is a random matrix chosen in the Orthogonal group $O(N)$ according to the Haar measure and $\b B$ is a given random matrix independent from $\b C$ and $\b O$ (see e.g.\ \cite{burda2005spectral, couillet2015kernel, bun2015rotational} for similar models). The framework investigated above corresponds to the case where $\b O \b B \b O^*$ is a Wishart matrix. Using the results of \cite{bun2015rotational}, we find for this general model that \eqref{eq:psi_SCM} still holds with $\zeta(z) = \cal S_{\b B}(z\stj(z) - 1)$
where $\cal S_{\b B}$ is the so-called Voiculescu's $\str$-transform of the $\b B$ matrix \cite{voiculescu1991limit}. If $\b B = \b{\cal W}$, then $\cal S_{\b B}(\omega) = 1/(1+q\omega)$. However, it seems difficult at this stage to obtain an explicit formula for the mean squared overlap \eqref{eq:mso} in this general case since the analytic structure of the $\cal S$-transform depends on the choice of $\b B$. 

\subsection{Additive noise}
\label{sec:additive_noise}

All the arguments that we use for the multiplicative noise model can be repeated nearly verbatim for the additive noise. In the
case of additive real symmetric Gaussian noise, referred to as the Gaussian Orthogonal Ensemble (GOE) in the literature, we have $\b S = \b C + \b W$ and $\st\S = \C + \st{\b W}$ with $\b W$ and $\st{\b W}$ two independent GOE matrices with variance $\sigma^2$ and $\st\sigma^2$. Each entry of the resolvent $(z - \S)^{-1}$ may also be approximated by a deterministic value in the high-dimensional regime \cite{knowles2014anisotropic,allez2014eigenvector,bun2015rotational}:
\begin{equation}
\label{eq:AILL_dGOE}
\E_{\cal P}\qb{(z - \b S)^{-1}_{kl}} \sim \pB{\zeta^a(z) - \b C}_{kl}^{-1}\,,
\end{equation}
where we defined \footnote{Here and henceforth, the superscript $a$ denotes the additive noise.}
\begin{equation}
	\zeta^a(z) \;\deq\; z - \sigma^2 \stj(z)\,.
\end{equation}
Once again, the asymptotic limit \eqref{eq:AILL_dGOE} holds for $\st{\S}$ by replacing $\zeta^a$, $\stj$ and $\sigma$ by $\st\zeta^a$, $\st\stj$ and $\st\sigma$. By performing the same computations as above, we obtain for the limiting value of $\psi_N$ :
\begin{equation}
\label{eq:psi_dGOE}
\psi^a(z, \st z) = \frac{\stj(z) - \st\stj (\st z)}{\st\zeta^a(\st z) - \zeta^a(z)}\,.
\end{equation}
As for Eq.\ \eqref{eq:psi_SCM}, Eq.\ \eqref{eq:psi_dGOE} depends only on \emph{a priori} observable quantities since it does not involve explicitly the unknown population matrix $\b C$. Consequently, we will obtain an observable expression for \eqref{eq:mso} using the inversion formula \eqref{eq:inversion_formula}. 

We can now turn on the computation of the mean squared overlap \eqref{eq:mso} in the additive Gaussian model. From \eqref{eq:psi_dGOE}, we find that
\begin{align}
	\lim_{\eta\to0}\qb{\psi^a(\lambda-\ii\eta,\st\lambda + \ii\eta) - \psi^a(\lambda-\ii\eta,\st\lambda - \ii\eta)}  \;=\; \nonumber \\
	\quad \frac{\stj_0\pB{\st \zeta^a - \ovl{\st \zeta^a}\,} + \zeta^a \pB{\ovl{\st \stj_0} - \st \stj_0} + \stj_0 \, \ovl{\st \zeta^a}  -  \ovl{\st \stj_0} \, \st \zeta^a }{\pB{\ovl{\st \zeta^a} - \zeta^a}\pB{{\st \zeta^a} - \zeta^a}} \nonumber
\end{align}
where we used the notation $\stj_0 \equiv \lim_{\eta\downarrow0^+}\stj$. Defining $\zeta^a_0 = \lim_{\eta \downarrow 0^+}\zeta^a(\lambda-\ii\eta) \equiv \zeta^a_R + \ii \zeta^a_I$ and performing similar algebraic manipulations as above (see Appendix \ref{app:mso_dGOE} for details), we eventually get 
\begin{align}
	\label{eq:mso_dGOE}
	\Phi^a_{\sigma, \tilde\sigma}(\lambda, \st\lambda) = \frac{ 2(\st\sigma^2 \lambda - \sigma^2 \st \lambda)(\zeta^a_R - \st\zeta^a_R) + (\sigma^2 - \st \sigma^2)\beta^a(\lambda,\st\lambda) }{\gamma^a(\lambda,\st\lambda) }
\end{align}
with $\gamma^a(\lambda,\st\lambda)$ given by the same expression as $\gamma(\lambda,\st\lambda)$ in Eq.\ \eqref{eq:mso_SCM_general_aux_var} with the substitutions $m_R \to \zeta^a_R$ and $m_I \to \zeta^a_I$ and 
\begin{align}
	\beta^a(\lambda, \st\lambda) & \;\deq\; \pb{\zeta^a_R(\lambda) - \zeta^a_I(\lambda)}\pb{\zeta^a_R(\lambda) + \zeta^a_I(\lambda)}  \nonumber\\
	& \qquad - \pb{\st\zeta^a_R(\st\lambda) - \st\zeta^a_I(\st\lambda)}\pb{\st\zeta^a_R(\st\lambda) - \st\zeta^a_I(\st\lambda)}\,.
\end{align}
We notice that the contribution of the term $\beta^a$ again vanishes when we suppose that both Gaussian noises have the same variance. As in the multiplicative case,  the self-overlap $\Phi^{a}(\lambda, \lambda)$ is reached by expanding \eqref{eq:mso_dGOE} in power of $\e$ with $\st\lambda = \lambda + \e$. This eventually yields by taking $\e = 0$:
\begin{equation}
	\label{eq:mso_GOE_bulk_same_q_eig}
	\Phi^a(\lambda,\lambda) = \frac{\sigma^2}{2} \frac{\partial_\lambda \zeta^a_R(\lambda)}{(\zeta^a_I(\lambda))^2 \absb{ \partial_\lambda \zeta^a_0(\lambda)}^2}\,.
\end{equation}

Similarly to the multiplicative case, the additive model can be generalized to $\b S = \b C + \b O \b B \b O^*$ with the same definitions for $\b B$ and $\b O$. In that case, the above result \eqref{eq:psi_dGOE} holds but now with $\xi(z) = z - \cal R_{\b B}(\stj(z))$ where $\rtr_{\b B}(z)$ is the $\rtr$-transform of the $\b B$ matrix \cite{voiculescu1991limit} -- which is simply equal to 
$\cal R_{\b B}(z)=\sigma^2 z$ when $\b B= \b W$ is a Gaussian random matrix, as considered above.

\begin{figure}[ht]
	\centering
   \includegraphics[scale = 0.4]{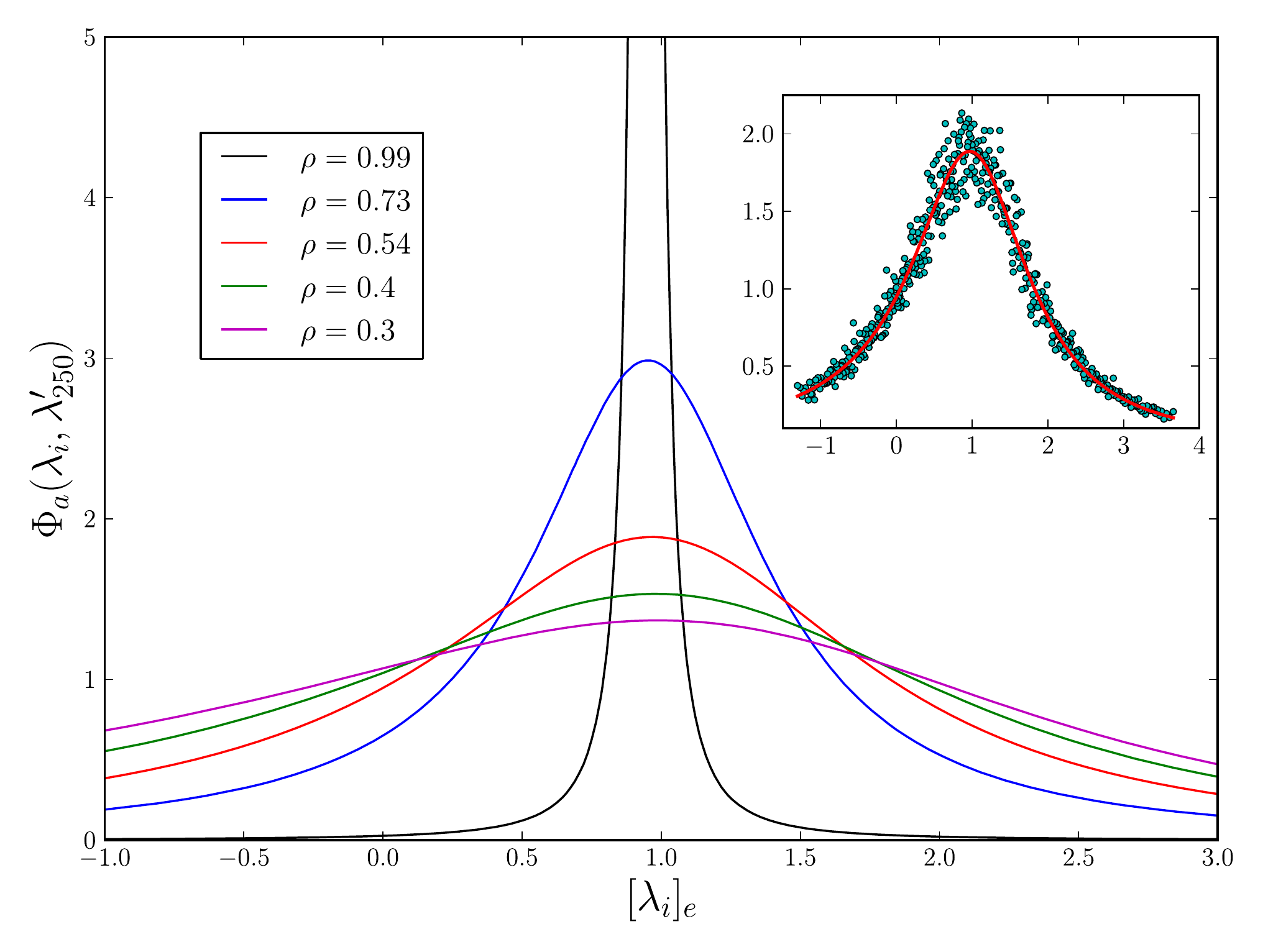} 
   \caption{Main Figure: Evaluation of the self overlap $\Phi_a(\lambda,\st{\lambda})$ for a fixed $\st{\lambda} \approx 0.95$ as a function of $\lambda$ for $N = 500$, $\sigma = 1$, and for different values of $\rho$. The population matrix $\C$ is given by a (white) Wishart matrix with parameter $T = 2N$. Inset: We compare the theoretical prediction $\Phi_a(\lambda,\st{\lambda} \approx 0.95)$ for a fixed $\rho = 0.54$ with synthetic data. 
   The empirical averages (blue points) are obtained from 100 independent realizations of $\b S$.} 
   \label{fig:mso_addition_different_rho_inset}
\end{figure}

Another interesting and important extension of the result \eqref{eq:mso_dGOE} is when the noises $\b W$, $\st{\b W}$ are correlated -- while the above calculations referred to independent noises. 
In the additive case, the trick is to realize that one can always write (in law) 
$\b W= \sqrt{\rho} \, \b W_0 + \sqrt{1 - \rho} \, \b W_1$ and $\st{\b W}= \sqrt{\rho} \, \b W_0 + \sqrt{1 - \rho} \, \b W_2$, where $\b W_1$, $\b W_2$ are now independent, as above. 
Since our formulas do not rely on the common matrix
$\b C$, it can therefore be replaced by $\b C + \sqrt{\rho} \b W_0$. Then, Eq.\ \eqref{eq:mso_GOE_bulk_same_q_eig} trivially holds with $\sigma^2$ replaced by $\sigma_1^2 - \sigma_1 \sigma_2 \rho$ with $\sigma_1$ and $\sigma_2$ the width of the noisy matrices $\b W_1$ and $\b W_2$ (see Appendix \ref{app:mso_dGOE_corr} for more details). Similarly, $\st{\sigma}^2$ is replaced by $\sigma_2^2 - \sigma_1 \sigma_2 \rho$. Note that in the case where the noises parameter are identical, $\sigma^2$ is simply 
multiplied by $1 - \rho$. The corresponding shape of $\Phi^a(\lambda,\lambda)$ for 
different values of $\rho$ and  $\sigma = \st{\sigma}$ is shown in Fig.\ \ref{fig:mso_addition_different_rho_inset}. We also provide in the inset a comparison with synthetic data for a fixed $\rho = 0.54$, $\sigma = 1$. The empirical average is taken over 200 realizations of the noise and again, the agreement is excellent. 

Considering correlated noises in the multiplicative model is of crucial importance since it describes the case of correlation matrices measured on overlapping periods, such that 
$\b S = \sqrt{\b C} \left[\b{\cal W}_0 + \b{\cal W}\right] \sqrt{\b C}$ and $\b S' = \sqrt{\b C} \left[\b{\cal W}_0 + \st{\Wishart}\right] \sqrt{\b C}$ (see Section \ref{sec:RIE} below). 
This case turns out to be more subtle and is the subject of ongoing investigations.

\subsection{Convolution formula}
\label{sec:convolution}

Before investigating some concrete applications of these formulas, let us end this theoretical section with an interesting interpretation of the above formalism. We first introduce 
the set of eigenvectors ${\bf v}_\mu$ of the pure matrix $\b C$, labeled by the eigenvalue $\mu$. We define the overlaps between the ${\bf u}$'s and the ${\bf v}$'s as 
$\sqrt{\Phi(\mu,\lambda)/N} \times \epsilon(\mu,\lambda)$, where $\Phi(\mu,\lambda)$ were explicitly computed in \cite{bun2015rotational} for a wide class of problems, 
and  $\epsilon(\mu,\lambda)$ are random variables of unit variance. Now, one can always decompose the ${\bf u}$'s as:
\begin{equation}
{\bf u}_\lambda = \frac{1}{\sqrt{N}} \int {\rm d}\mu \varrho_C(\mu) \sqrt{\Phi(\mu,\lambda)} \epsilon(\mu,\lambda) {\bf v}_\mu,
\end{equation}
where $\varrho_C$ is the spectral density of $\b C$. Using the orthonormality of the ${\bf v}$'s, one then finds:
\begin{equation*}
\scalar{{\bf u}_\lambda}{{\bf u}'_{\lambda'}} = \frac1N \int {\rm d}\mu \varrho_C(\mu) \sqrt{\Phi(\mu,\lambda)\Phi(\mu,\lambda')} \, \epsilon(\mu,\lambda)\epsilon(\mu,\lambda').
\end{equation*}
If we square this last expression and average over the noise, and make an ``ergodic hypothesis" \cite{deutsch1991quantum} according to which all signs $\epsilon(\mu,\lambda)$ are 
in fact independent from one another, one finds the following, rather intuitive convolution result for squared overlaps:
\begin{equation}
\label{eq:overlap_convolution}
\Phi(\lambda,\lambda')  = 
\int {\rm d}\mu \varrho_C(\mu) \Phi(\mu,\lambda)\, \Phi(\mu,\lambda').
\end{equation}
It turns out that this expression is completely general and exactly equivalent to Eqs.\ \eqref{eq:mso_SCM_general} and \eqref{eq:mso_dGOE} in the corresponding cases. However, whereas this 
expression still contains some explicit dependence on the structure of the pure matrix $\b C$, it has disappeared in Eqs.\ \eqref{eq:mso_SCM_general} and \eqref{eq:mso_dGOE}. Nevertheless, this second interpretation will be useful in order to obtain an efficient way to estimate $\C$ from large noisy matrices.

\section{Application}

Eqs.\ \eqref{eq:mso_SCM_general} and \eqref{eq:mso_dGOE} are exact in the high-dimensional limit (HDL) and are the main new results of this work. Note that from Ref. \cite{knowles2014anisotropic}, we expect these results to hold with fluctuations of order $N^{-1/2}$ but a more rigorous analysis of the subleading terms can be useful for practical purposes. We leave this question for future works. Throughout this section, we shall focus on the case of sample covariance/correlation matrices but most of the results that will follow can be easily transposed to the additive noise. We emphasize that in the special case of sample correlation matrices, the HDL is defined as
\begin{equation}
	\label{eq:HDL_SCM}
	N\,, T\,, \st T\, \to \infty \quad{\text{with}}\quad q \;=\; \cal O(1)\,,\; \st{q} \;=\; \cal O(1)\,,
\end{equation}
where $T$ is the sample size. We will assume throughout this section that the variance of each variable can be estimated independently with great accuracy in the HDL so that we will not distinguish further covariances and correlations henceforth. 

The first application concerns a stability test for the eigenvectors of large correlation matrices. More precisely, we investigate whether or not the mean squared overlap between the eigenvectors of two correlation matrices measured with nearby non-overlapping samples is entirely explain by measurement noise. Differently said, we test the hypothesis that the dynamics of the eigenvectors is captured by the sample correlation matrix model. The second application involves the convolution formula. In particular, we link our results with the theory of RIEs that provides significant improvement over classical sample estimates in the HDL (see \cite{bun2017cleaning} for a recent review).

\subsection{Eigenvectors stability}

\begin{figure}[ht]
	\centering
   \includegraphics[scale = 0.38]{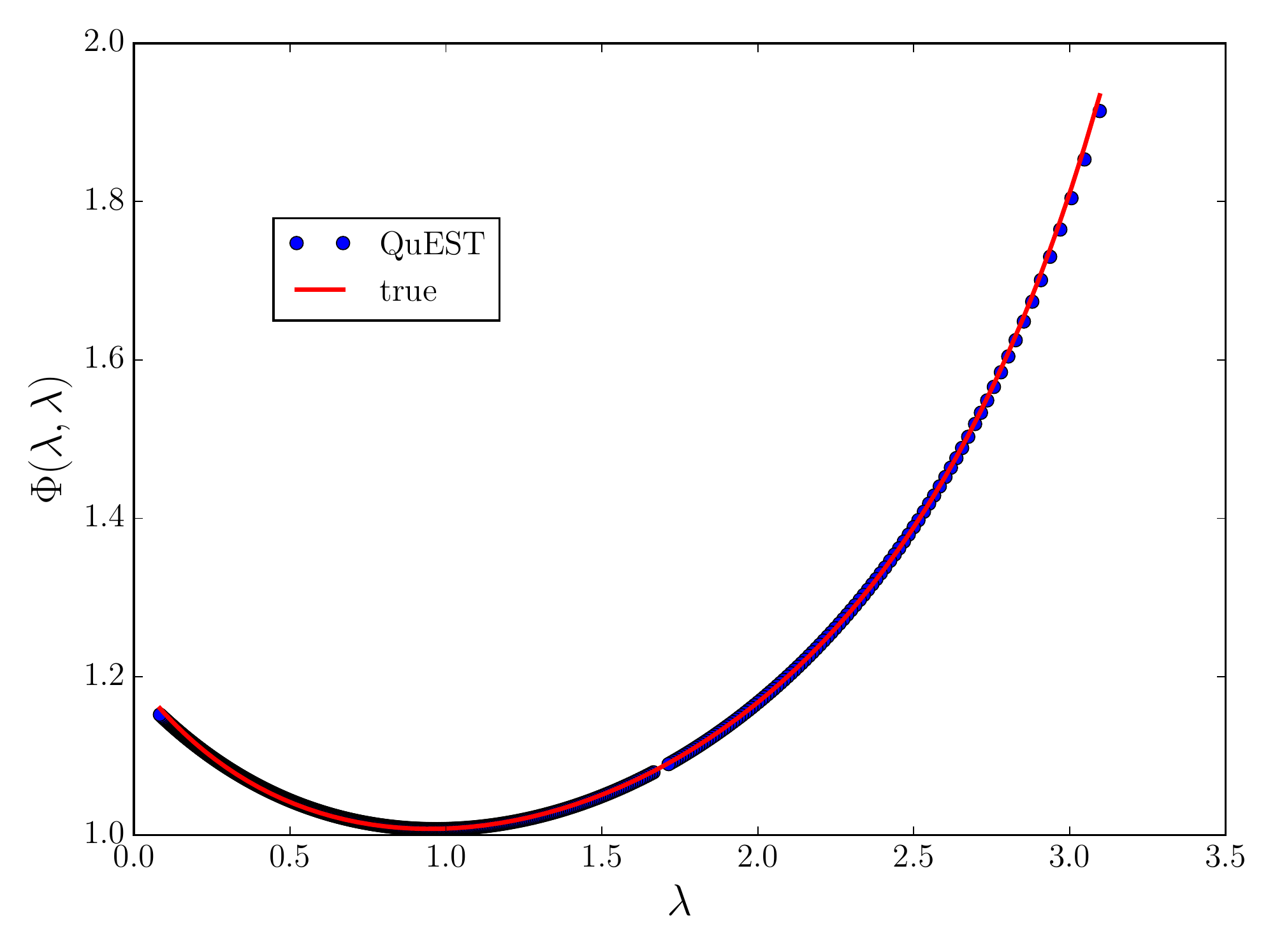} 
   \caption{Evaluation of the self-overlap for an Inverse-Wishart population matrix with $N = 500$ and $\kappa = 10$ using the formula \eqref{eq:mso_same_q_eig_invW_exp}. The red plain line corresponds to the case where we know exactly the true eigenvalues while the blue points are obtained from the estimated population eigenvalues using the QuEST algorithm (see  \cite{ledoit2016numerical}). The empirical average is plotted by the green diamond-line and is obtained from \eqref{eq:mso_empirical_proc} over 50 realizations of $\Wishart$. }
\label{fig:mso_invwishart}
\end{figure}

The first application deals with the stability of the eigenvectors in the case of two non-overlapping adjacent samples. In order to give more insights, we begin with a theoretical example where the true correlation matrix $\b C$ is an inverse Wishart matrix
of parameter $\kappa \in (0,\infty)$, that corresponds to $1/q$ for Wishart matrices (see \cite{bun2015rotational} for details). In that case, the function $m(z)$ can be explicitly computed. 
This finally leads to:
\begin{equation}
	\label{eq:mso_same_q_eig_invW}	
	\Phi(\lambda,\lambda) =  \frac{\upsilon(\lambda + 2q\kappa)^2}{2 q \kappa \pb{2\lambda(\upsilon+\kappa) - \lambda^2\kappa +\kappa(2q\upsilon-1)}}
\end{equation}
with  $\upsilon := 1+q \kappa$ and $\lambda$ is within the interval $[\lambda^{-},\lambda^{+}]$, where the edges are given by $\lambda^{\pm} =  
\kappa^{-1} \qB{ \upsilon + \kappa \pm \sqrt{(2\kappa+1)(2q\kappa+1)}}$. An interesting limit corresponds to $\kappa \to \infty$, where $\b C$ tends to 
the identity matrix, and the overlaps are expected to become all equal to $1/N$. Indeed one finds, for a fixed $q$: 
\begin{equation}
	\label{eq:mso_same_q_eig_invW_exp}	
	\Phi(\lambda,\lambda') \underset{\kappa\to\infty}{\sim} \qBB{1+\frac{(\lambda - 1)(\lambda'-1)}{2q^2\kappa} + \cal O\pB{\frac{1}{\kappa^2}}},
\end{equation}
which is in fact {\it universal} in this limit, provided the eigenvalue spectrum of $\b C$ has a variance given by $(2\kappa)^{-1} \to 0$ 
\footnote{The analysis for the additive case leads to a very similar result. More precisely, taking $\b C = I_N + \b W_0$ with $\b W_0$ a GOE 
(independent from $\b W$ and $\b W'$) of variance $\sigma_0^2 \to 0$,
one finds that $\Phi_a(\lambda,\lambda')$ is given by exactly the same formula \eqref{eq:mso_same_q_eig_invW_exp} with the substitution $2 q^2 \kappa \to \sigma^4/\sigma_0^2$.}.
This formula is interesting insofar as it allows one to estimate the width of the eigenvalue distribution of $\b C$, even when it is close to the identity matrix, 
i.e. $\kappa \gg 1$. One could think of directly using information on the empirical spectrum, for example the Mar\v{c}enko-Pastur prediction $\tr\C^{-1} = (1-q) \tr\S^{-1}$, that in principle allows one extract the parameter $\kappa$ through $1 + (2\kappa)^{-1} = (1-q) \tr\S^{-1}/N$. However, this second method is numerically unstable and very imprecise when $\kappa \gg 1$ and finite $N$ (for one thing, the RHS can be negative, which would lead to a negative variance). Our formula based on 
overlaps avoid these difficulties. As an illustration, we check the validity of Eq.\ \eqref{eq:mso_same_q_eig_invW} in Figure \ref{fig:mso_invwishart} with $\kappa = 10$, $N = 500$ and $q=0.5$. More precisely, we determine the empirical average overlap as follows: we consider 50 
independent realization of the Wishart noise $\b{\cal W}$. For each pair of samples we compute a smoothed overlap as:
\begin{equation}
	\label{eq:mso_empirical_proc}
	\qb{\scalar{\b u_i}{ {\st{\b u}_i}}^2} = \frac{1}{Z_i} \sum_{j = 1}^{N} \frac{\scalar{\b u_i}{{\st{\b u}}_j}^2}{(\lambda_i - \lambda'_j)^2 + \eta^2},
\end{equation}
with $Z_i = \sum_{k=1}^{N} ( (\lambda_i - \lambda'_k)^2 + \eta^2)^{-1}$ the normalization constant and $\eta$ the width of the Cauchy kernel, that we 
choose to be $N^{-1/2}$ in such a way that $N^{-1} \ll \eta \ll 1$. We then average this quantity over all pairs for a given value of $i$ to obtain 
$[ \scalar{\b u_i}{ {\st{\b u}_i}}^2 ]_e$, and plot the 
resulting quantity as a function of the average eigenvalue position $[\lambda_i]_e$. We observe that the agreement with Eq.\ \eqref{eq:mso_same_q_eig_invW} is excellent, even when the true underlying matrix $\b C$ is close to the identity matrix. Note that the empirical estimate Eq.\ \eqref{eq:mso_empirical_proc} is universal, i.e.\ independent of the underlying structure of $\C$. 

For a general and arbitrary population matrix $\C$, evaluating Eq.\ \eqref{eq:mso_SCM_general} -- or Eq.\ \eqref{eq:mso_dGOE} -- is rather difficult because of finite size effects, especially for the multiplicative case. Indeed, when we consider multiplicative noises, the eigenvalues of $\S$ are confined to stay positive meaning the presence of a \emph{hard} wall at the origin \footnote{Note that this problem does not appear in the additive case, so one can safely use concentration results from \cite{knowles2014anisotropic}.}. Consequently, the use of local laws  to estimate the Stieltjes transform $\stj(z)$, as in \cite{knowles2014anisotropic}, often leads to noisy results for very small eigenvalues. Hence, the determination of Eq.\ \eqref{eq:mso_SCM_general} from real data is rather difficult and one has to resort to numerical regularization schemes to do so. In the specific case of sample correlation matrix, one possible solution is to invert the celebrated Mar{\v c}enko-Pastur equation \cite{karoui2008spectrum,ledoit2016numerical} to infer the eigenvalues of the population matrix $\C$. Once this is done, one can evaluate the Stieltjes transform $\stj(z)$ with high precision even near the origin. In the following, we shall use the so-called QuEST numerical scheme of \cite{ledoit2016numerical} to obtain these \emph{pure} eigenvalues. We plot in Figure \ref{fig:mso_invwishart} the results obtained when using the estimated population eigenvalues from the QuEST algorithm (blue points) and note that the agreement is quite remarkable.

Now that we have an estimate of the population eigenvalues, we can investigate an application to real data. Here, we study the case of US stock market but the results below can be extended to other region \cite{bun2017cleaning}. The difficulty when dealing with real data is to measure the empirical mean squared overlaps \eqref{eq:mso_empirical_proc} between two non-overlapping correlation matrices $\S$ and $\st{\S}$ as in Eq. \eqref{eq:mso_empirical_proc} because we may not have enough data points to evaluate accurately an average over the noise as required in Eq.\ \eqref{eq:mso}. To circumvent this problem, we use a Bootstrap procedure to increase the size of the data \footnote{This technique is especially useful in machine learning and we refer the reader to e.g. \cite[Section 7.11]{friedman2001elements} for a more detailed explanation.}: we take a total period of 2400 business days from 2004 to 2013 for the $N = 300$ most liquids assets of the S\&P 500 index that we split into two non-overlapping subsets of same size of 1200 days, corresponding to 2004 to 2008 and 2008 to 2013. We restrict to $N = 300$ stocks such that all of them are present throughout the whole period from 2004 to 2013.  Then, for each subset and each 
Bootstrap sample $b \in \{1,\dots, B\}$, we select randomly $T=600$ distinct days to construct two ``independent'' sample correlation matrices 
$\S_b$ and $\st{\S}_{b}$, with $q= \st q = N/T=0.5$. We then compute the empirical mean squared overlap \eqref{eq:mso} and also the theoretical limit \eqref{eq:mso_SCM_general} -- using the QuEST algorithm -- from these $B$ bootstrap data-sets. 

\begin{figure}[ht]
	\centering
   \includegraphics[scale = 0.4]{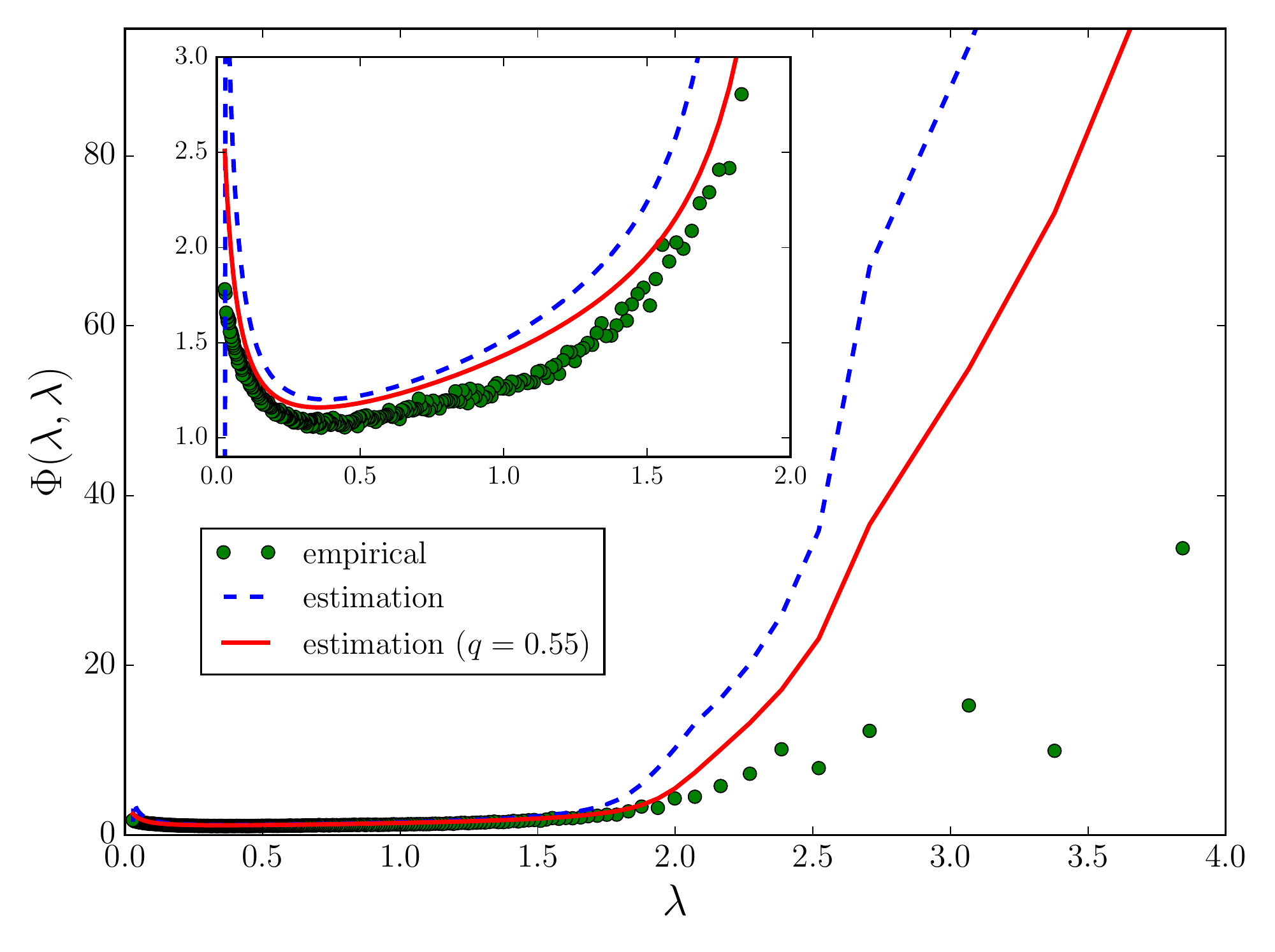}
   \caption{Evaluation of the self-overlap $\Phi(\lambda,\lambda)$ as a function of the sample eigenvalues $\lambda$ using the $N=300$ most liquid US equities from 2004 to 2013. 
   We split the data into two non-overlapping period with same sample size 1200 business days. For each period, we randomly select $T=600$ days and we repeat $B=100$ bootstraps of the original data. The empirical self-overlap is computed using Eq.\ \eqref{eq:mso_empirical_proc} over these 100 bootstraps (green points) and the limiting formula \eqref{eq:mso_SCM_bulk_same_q_eig} is estimated using the QuEST algorithm with $q=0.5$ (blue dashed line). We also provide the estimation we get using the same effective observation ratio $q_{\text{eff}} = 0.55$ that accounts for correlation/heavy tail effects \cite{bun2017cleaning}. Inset: focus in the bulk of eigenvalues.}
   \label{fig:mso_SPX}
\end{figure}

For our simulations, we set $B = 100$ and plot in Figure \ref{fig:mso_SPX} the resulting estimation of Eq.\ \eqref{eq:mso} we get from the QuEST algorithm (blue dashed line) and the empirical 
bootstrap estimate \eqref{eq:mso_empirical_proc} (green points) using US stocks. We also perform the estimation with an effective observation ratio $q_{\text{eff}} = 0.55$ (red plain line) as advocated in \cite{bun2016beautiful} for the S\&P 500 index, to account for correlation or heavy tail effects.

It is clear from Figure \ref{fig:mso_SPX} that the eigenvectors associated to large eigenvalues are not well described by the theory: we notice a discrepancy between the (estimated) theoretical curve and the empirical one even after accounting for an effective ratio $q_{\text{eff}}$. The difference is even worse for the market mode (not shown). This is presumably related to the fact that the largest eigenvectors are expected to genuinely evolve with time, as already 
argued in \cite{allez2012eigenvector}. Note also the gap at the left edge between the theoretical and empirical prediction in the inset of Figure \ref{fig:mso_SPX} that is partly corrected with $q_{\text{eff}}$. This suggests that one can still improve the Mar{\v c}enko-Pastur framework by adding e.g. autocorrelation or heavy tailed entries which allows one to widen the LSD of $\S$ (see e.g.\ \cite{burda2005spectral,bartz2014covariance} for autocorrelation and \cite{biroli2007student,burda2004free,el2009concentration,couillet2015random} for heavy tailed entries). Finally, all these remarks hold for other markets as well \cite{bun2017cleaning}.

\subsection{Rotational invariant estimator}
\label{sec:RIE}

Aside from the statistics of eigenvectors, the theoretical framework presented is actually very useful for the estimation of $\C$ from large noisy matrices in the specific class of rotational invariant estimators (see \cite{bun2017cleaning} and references therein for a recent review on this topic). For this class of estimators, the crux is to find an observable way to estimate the \emph{oracle} eigenvalues
\begin{equation}
	\label{def:oracle}
	\xi_{i} \;\equiv\; \xi_i(q) \;\deq\; \scalar{\b u_i}{\C \b u_i}\,,
\end{equation}
where the $\b u_i$ are the eigenvectors of $\S := \sqrt{\C} {\Wishart} \sqrt{\C}$, and we omit their dependence on $q$ in the RHS of the last equation. There exist several way to approximate this estimator in the HDL: using the anisotropic local law \cite{bun2017cleaning}, numerical inversion of Mar{\v c}enko-Pastur equation \cite{ledoit2016numerical} and the so-called cross-validation (CV) estimator of \cite{bartz2016cross}. The remarkable feature of all these methods is that the resulting estimator only depends on observable quantities, while it is clearly not the case for \eqref{def:oracle}. Even if the first two techniques are interesting on their own, we will rather focus on the last one as it turns to be related to the convolution formula derived in Section \ref{sec:convolution}. 

From now on, we consider the multiplicative but the following arguments can easily be generalized for the additive case. Let us consider the quantity
\begin{equation}
	\nu_i \;\equiv\; \nu_i(q) \;\deq\; \scalar{\b u_i}{\st{\b S} \b u_i}\,,
\end{equation}
where we recall that $\b u_i$ is independent from $\st\S := \sqrt{\C} \st{\Wishart} \sqrt{\C}$. We again assume that we are in the regime \eqref{eq:HDL_SCM} and we thus infer from \eqref{eq:overlap_convolution} that
\begin{align}
	\label{eq:oracle_convolution_tmp}
	\nu_i & \;\sim \; \int \tilde \varrho(\tilde\lambda) \Phi(\lambda, \tilde\lambda) \tilde\lambda\, \dd\tilde\lambda \nonumber \\
	& \;=\; \int \tilde \varrho(\tilde\lambda) \qBB{ \int \varrho_{\C}(\mu) \Phi(\lambda,\mu)\Phi(\tilde\lambda, \mu) \dd\mu} \tilde\lambda \,\dd\tilde\lambda  \nonumber \\
	& \;=\; \int \varrho_{\C}(\mu) \Phi(\lambda,\mu) \qBB{ \int \tilde \varrho(\tilde\lambda)  \Phi(\tilde\lambda, \mu) \tilde\lambda \,\dd\tilde\lambda } \, \dd\mu\,.
\end{align}
The term in the bracket can be simplified by using the very definition of $\st{\b S}$: 
\begin{align*}
	\int \tilde \varrho(\tilde\lambda)  \Phi(\tilde\lambda, \mu_j) \tilde\lambda \,\dd\tilde\lambda & \;\approx\;  \scalar{\b v_j}{\tilde{\b S} \b v_j} \nonumber \\
	& \;=\; \scalar{\C^{1/2} \b v_j}{ \tilde{\b{\cal W}} \C^{1/2} \b v_j}\,,
\end{align*}
and we rewrite this last line thanks to the eigenvalue equation as 
\begin{align*}
	\int \tilde \varrho(\tilde\lambda)  \Phi(\tilde\lambda, \mu_j) \tilde\lambda \dd\tilde\lambda & \;\approx\;  \mu_j \scalar{\b v_j}{ \tilde{\b{\cal W}} \b v_j} \nonumber \\
	& \;=\; \mu_j \sum_{k=1}^{N} \omega_k \mathbb{E}\qb{\scalar{\b w_k}{\b v_j}^2}\,,
\end{align*}
with $\omega_k$ the $k$-th eigenvalue of the white Wishart matrix $\tilde{\b{\cal W}}$ and $\b w_k$ the corresponding eigenvector. Finally, we invoke that $ \mathbb{E}[\scalar{\b w_k}{\b v_j}^2] = N^{-1}$ for all $j \in \qq{1,N}$ to conclude that in the HDL,
\begin{equation}
	\sum_{k=1}^{N} \omega_k \scalar{\b w_k}{\b v_j}^2 \;=\; 1\,
\end{equation}
and we therefore have
\begin{align}
	\int \tilde \varrho(\tilde\lambda)  \Phi(\tilde\lambda, \mu_j) \tilde\lambda \,\dd\tilde\lambda & \;\approx\;   \mu_j.
\end{align}
Plugging this last equation into \eqref{eq:oracle_convolution_tmp}, we obtain for any
$\st{q} = \cal O(1)$ the following result:
\begin{equation}
	\label{eq:oracle_convolution}
	\nu_i(q) \;\sim\;  \int \varrho_{\C}(\mu) \,\Phi(\lambda,\mu)\, \mu \,\dd\mu \;\equiv\; \xi_i(q)\,,
\end{equation}
where the last equivalence comes from the definition \eqref{def:oracle} of the oracle estimator in the continuous limit. 

This result is very interesting and indicates  that one can approximate the oracle estimator \eqref{def:oracle} by considering the quadratic form between the eigenvectors of a given realization of $\C$ -- say ${\b S}$ -- and another realization of $\C$ -- say $\st{\b S}$ -- even if the latter is characterized by a different value of the quality ratio $\st q \neq q$.

\begin{figure}[ht]
	\centering
   \includegraphics[scale = 0.4]{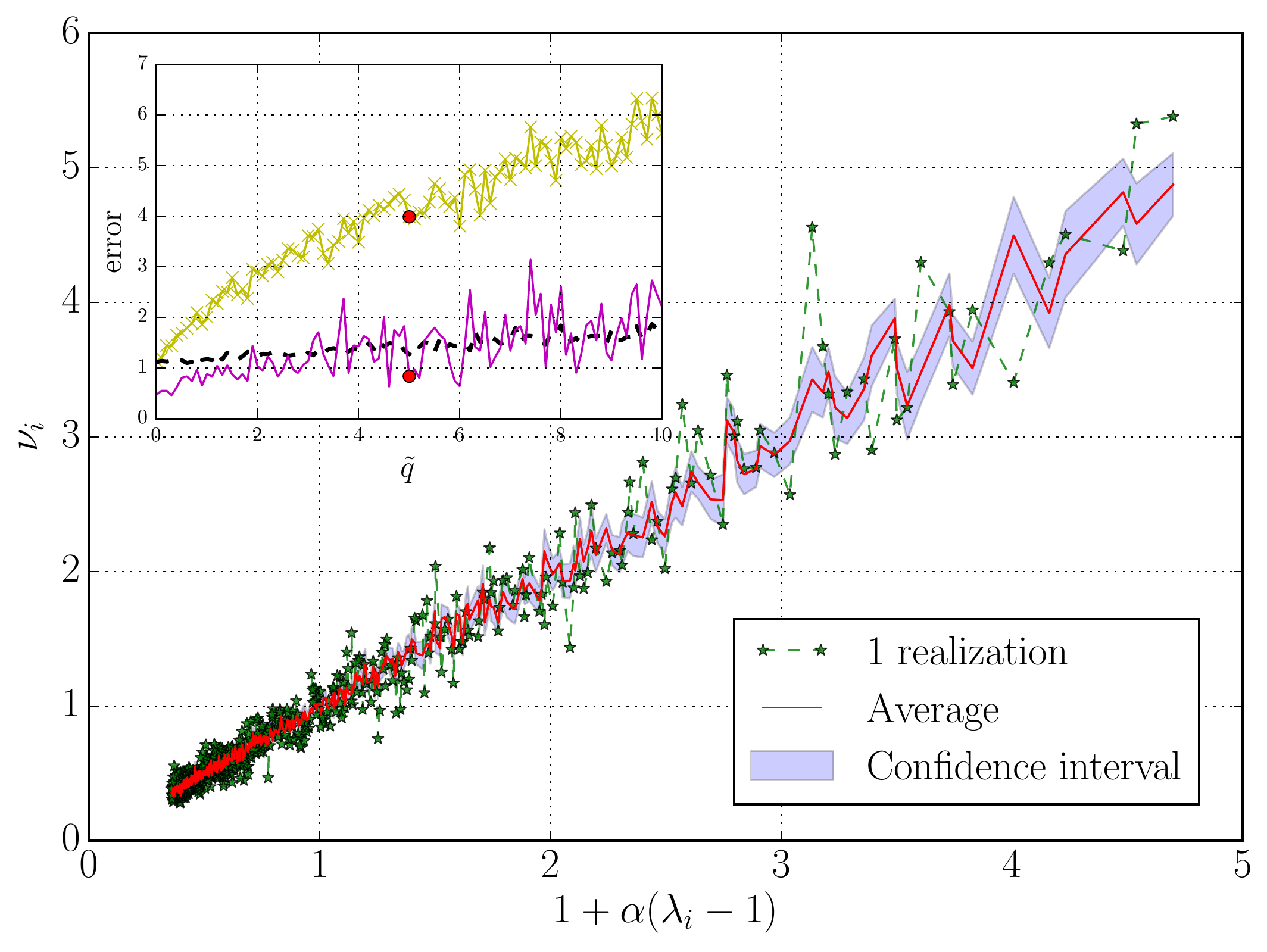} 
   \caption{Main figure: Evaluation of Eq.\ \eqref{eq:oracle_convolution} when $\C$ is a $500 \times 500$ Inverse-Wishart matrix with parameter $\kappa = 0.5$. The first noisy matrix $\S$, which is the one we wish to denoise, is drawn from a Wishart distribution with parameter $q = 0.5$. The second noisy matrix $\st{\S}$ is also a Wishart matrix but with parameter $\st{q} = 5$. The x-axis is given by the true asymptotic value, namely the linear shrinkage estimator with intensity $\alpha = 1/(1 + 2q\kappa)$. The y-axis are the eigenvalues obtained from \eqref{eq:oracle_convolution}. The dashed star line is the result obtained from one sample, the plain red line is the average results from 20 independent realizations of $\st{\S}$ and the blue shaded are gives the confidence interval. Inset: Squared error as a function of $\st{q}$ where we kept $\S$ fixed with the same $N$ and $T$. The yellow crossed-line corresponds to the error over a single realization of $\st{\S}$ for Eq.\ \eqref{eq:oracle_convolution} and the purple plain line to its sorted version. The dashed black line corresponds to the average value of Eq.\ \eqref{eq:oracle_convolution} over 20 realizations of $\st{\S}$. The red point gives the error for $\st{q} = 5$, i.e. the sample shown in the main figure.}
\label{fig:oracle_convolution_linear_shrinkage}
\end{figure}

To illustrate this last point, let us consider an Inverse-Wishart matrix with parameter $\kappa = 0.5$ as the population correlation matrix of size $N = 500$. Both noisy matrices are drawn from a multivariate Gaussian distribution but with different parameters: the first noisy matrix $\S$ is computed using $T = 1000$ while the second one $\st{\S}$ corresponds to
$\st{T} = 100$. With this prior, the oracle estimator is given in the HDL by the linear shrinkage
with intensity $\alpha = 1/(1 + 2q\kappa)$ and $q = N/T$. In Figure \ref{fig:oracle_convolution_linear_shrinkage}, we plot the prediction obtained from \eqref{eq:oracle_convolution} with $\S$ fixed
and a single realization of $\st{\S}$ (dotted star line) and we see that although noisy, the prediction is already fairly accurate. We also plot in the same figure the average value of \eqref{eq:oracle_convolution} over 20 independent realizations of $\st{\S}$ (plain red line) and we observe that the agreement with limiting value (given by the line $y=x$ in Figure \ref{fig:oracle_convolution_linear_shrinkage}) is excellent, with only very small fluctuations (see the confidence interval given by the blue shaded area). 

Quite surprisingly, we can  significantly improve the accuracy of the estimation by applying an \emph{ad-hoc} regularization even for a single realization of $\st{\S}$. More specifically, we see from Figure \ref{fig:oracle_convolution_linear_shrinkage} that the prediction \eqref{eq:oracle_convolution} does not necessarily preserve the order of the eigenvalues due to the finite size of the sample.  However, observing a non-monotonic cleaning scheme may be an unwanted feature within a rotational invariant assumption. Indeed, there is no reason \emph{a priori} to expect that it is optimal to modify the order of the eigenvalues, that is to say, the variance associated to the principal components. 

There are several ways to regularize the estimation obtained from \eqref{fig:oracle_convolution_linear_shrinkage}. We can either sort the cleaned eigenvalues \cite{bun2017cleaning} or perform an \emph{isotonic} regression \cite{bartz2016cross}. We provide an illustration of the sorting regularization in the inset of Figure \ref{fig:oracle_convolution_linear_shrinkage} over a single realization of $\st{\S}$ (purple plain line). The improvement over \eqref{eq:oracle_convolution} (yellow crossed-line) in terms of squared error is significant. Moreover, even if the estimation becomes quite noisy for large values of $\st{q}$, we notice that the quality of estimation is still on par with the average value over 20 realizations of $\st{\S}$ (black dashed line), which is quite remarkable.We also want to emphasize that in all cases, the error we obtain
is always less than 9, which is the error we get when keeping the
sample eigenvalues of $\S$ (see Inset of Fig. \ref{fig:oracle_convolution_linear_shrinkage}).

A takeaway from Figure \ref{fig:oracle_convolution_linear_shrinkage} is that we can indeed use the result \eqref{eq:oracle_convolution} even when $\st{q} \gg q$. Hence, this gives a simple way to check the quality of the estimation by comparing the in-sample result (i.e. the estimator we obtain using the information of $\S$) and the out-of-sample result (obtain with the information of $\st{\S}$). It therefore demonstrates the validity of the out-of-sample test of \cite{bun2017cleaning} for assessing the quality of \emph{empirical} optimal RIE using financial data. The second finding is that it is possible to estimate quite accurately the optimal oracle for a fixed value of $q$ by using a relatively small amount of independent data from a \emph{single} realization. 

\subsection{Cross-validation estimator}

Another interesting discussion is the comparison of the above results with the cross-validation (CV) estimator proposed in \cite{bartz2016cross}. Suppose that we want to estimate \eqref{def:oracle} out of $T$ independent samples that we split into $K$ non-overlapping sets whose indices are denoted by $\{\cal I_{\varsigma}\}_{\varsigma = 1}^{K}$. The CV estimator then reads
\begin{equation}
	\label{def:CV_estimator}
	\nu_i^{\text{cv}}(q_\varsigma) \;\deq\; \frac{1}{K} \sum_{\varsigma=1}^{K} \sum_{t \in \cal I_{\varsigma}} \scalarB{\b u_i^{(\varsigma)}}{\frac{\b x_t \b x_t^*}{\absb{\cal I_{\varsigma}}} \b u_i^{(\varsigma)}}\,,
\end{equation}
where each set $\cal I_{\varsigma}$ has equal size such that $K |\mathcal{I}_{\varsigma}| = T$, $\b {u}_i^{(\varsigma)}$ is the eigenvector associated to the $i$-th eigenvalue obtained from the sample correlation matrix in which we removed all the observations belonging to the set $\cal I_{\varsigma}$ for a fixed $\varsigma \in \qq{1,K}$ and $q_\varsigma \deq N/(T - \abs{\cal I_{\varsigma}})$ is the corresponding observation ratio. If we denote by $\st{\S}_{\varsigma}$ the sample correlation matrix associated to the observations of the set $\cal I_{\varsigma}$, then we can rewrite \eqref{def:CV_estimator} as 
\begin{equation}
	\label{eq:CV_estimator_2}
	\nu_i^{\text{cv}} \;=\; \frac1K \sum_{\varsigma=1}^{K}  \scalarB{\b u_i^{(\varsigma)}}{\st{\S}_{\varsigma} \b u_i^{(\varsigma)}}\,,
\end{equation}
which can be thought as an average version of the estimator \eqref{eq:oracle_convolution}. 

The crucial difference resides in the observation ratios $q_\varsigma$  associated to the eigenvector $\b u_i^{(\varsigma)}$ and $\st{q}_\varsigma \deq N/\abs{\cal I_{\varsigma}}$ associated to the matrix $\st{\S}_\varsigma$ . Hence, we clearly have a trade-off in the choice of the cardinality of the $\cal I_{\varsigma}$: choosing $\abs{\cal I_{\varsigma}}$ too large implies that $q_\varsigma$ deviates strongly from $q \deq N/T$ while a too small cardinality leads to the noisy limit $\st{q}_\varsigma \sim N$ in which the convergence \eqref{eq:oracle_convolution} becomes dubious. We therefore understand from this rewriting why the \emph{leave-one-out} case, i.e. $\abs{\cal I_\varsigma} = 1$, will not return a reliable estimate of the optimal value \eqref{def:oracle} even after averaging since $\st{q}_\varsigma = N$. For a suitable value of the cardinality $\abs{\cal I_\varsigma}$, one can expect that $\st{q}_\varsigma$ is small enough in order to be in the regime of Eq.\ \eqref{eq:oracle_convolution} and, more importantly, that $\b u_{i}^{(\varsigma)}$ is not too far from $\b u_i(q)$. In that case, provided that the samples are independent to each other, we shall have 
\begin{equation}
	\label{eq:CV_to_oracle}
	\nu_i^{\text{cv}}(q_\varsigma) \sim \nu_i(q)\,.
\end{equation}

\begin{figure}[ht]
	\centering
   \includegraphics[scale = 0.4]{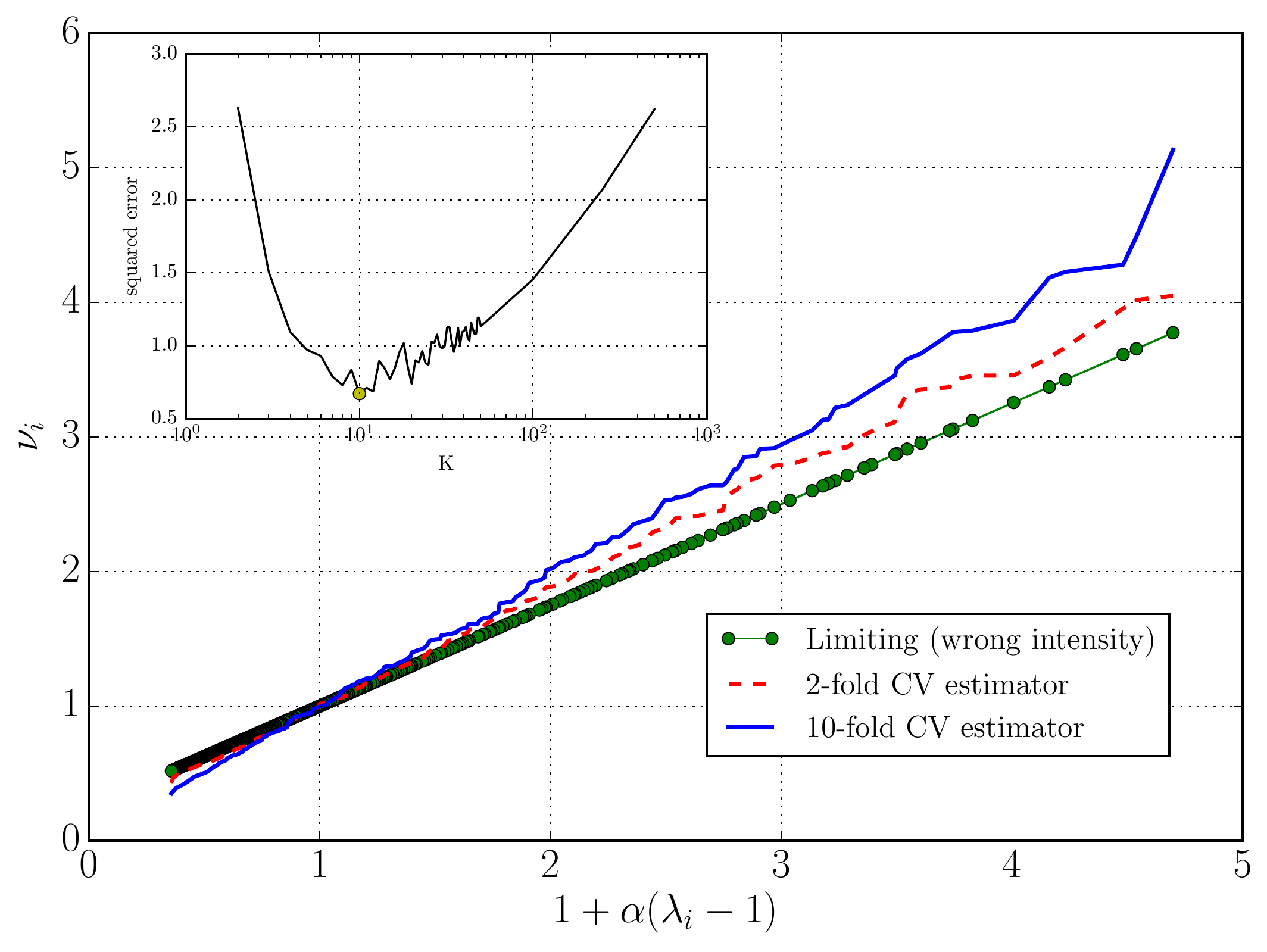} 
   \caption{Main figure: Evaluation of Eq.\ \eqref{def:CV_estimator} using the same experience than in Fig.\ \ref{fig:oracle_convolution_linear_shrinkage}. The x-axis represents to the optimal cleaning (with $\alpha = 1/(1 + 2q\kappa)$) while the y-axis are the eigenvalues obtained from \eqref{def:CV_estimator} for a single realization of $\S$. The blue plain line is the sorted 10-fold CV estimator ($\abs{\cal I_\varsigma} = 100$) and the red dashed line corresponds to the sorted $2$-fold case ($\abs{\cal I_\varsigma} = 500$). We also provide the linear shrinkage with a wrong intensity $\alpha = 1/(1+2q_\varsigma \kappa)$ where $q_\varsigma = 1$ corresponds to the case $\abs{\cal I_\varsigma} = 500$. Inset: We plot the squared error of \eqref{def:CV_estimator} for the same realization of $\S$ as a function of $K$. We see that the minimum is attained at $K=10$ (yellow point)}
\label{fig:CV_estimator}
\end{figure}

We provide a numerical check of this result in Figure \ref{fig:CV_estimator} in the case of a sorted 10-fold CV estimate (blue plain line) where we reconsider the same configuration of Figure \ref{fig:oracle_convolution_linear_shrinkage}. Note that a $10$-fold CV leads to a ``test'' set of size $\abs{\cal I_\varsigma} = 100$ meaning that $q_\varsigma \approx 0.55$ which is not that far from the true value of $q$. We see that the agreement with the optimal value is excellent showing that for this specific choice of $\abs{\cal I_\varsigma}$, the convergence \eqref{eq:CV_to_oracle} holds. In order to illustrate the trade-off discussed above, we also plot the $2$-fold CV estimator for which $q_\varsigma = 1$. In that case, the result we obtain from \eqref{def:CV_estimator} rather coincides with the linear shrinkage with an intensity of $\alpha = 1/(1+2q_\varsigma \kappa)$. This demonstrates that the choice of $K$ in \eqref{def:CV_estimator} is crucial in order to obtain a consistent estimation of \eqref{def:oracle} (see the inset of Figure \ref{fig:CV_estimator}). 

From a practical viewpoint, the estimator \eqref{eq:CV_estimator_2} is quite appealing since it can be generalized to more general classes of random processes. It provides a simple tool to approximate the oracle estimator \eqref{def:oracle} with great accuracy in the high-dimensional regime beyond the sample correlation matrix model. From a theoretical perspective, we see that there is still some work to be done in order to understand the convergence expressed by \eqref{eq:CV_to_oracle}. Indeed, the relationship between $\b u_i^{(\varsigma)}$ and $\b u_i$, which corresponds to the overlapping/correlated case discussed above in the additive case, should provide insights to establish the convergence result as a function of the cardinality of $\abs{\cal I_\varsigma}$. Moreover, the study of subleading terms in Eq.\ \eqref{eq:oracle_convolution} can be of particular interest to understand the phase transition that occurs depending on the value of $q_\varsigma$ and $\st{q}_\varsigma$.

\section{Conclusion}

In summary, we have provided general, exact formulas for the overlaps between the eigenvectors of large correlated random matrices, with additive or 
multiplicative noise. Remarkably, these results do not require the knowledge of the underlying ``pure'' matrix and have a broad range of applications in different contexts. 
We showed that the cross-sample eigenvector overlaps provide unprecedented information about the structure of the true eigenvalue spectrum, much beyond that contained in the empirical spectrum itself. For 
example, the width of the bulk of the spectrum of the true underlying correlation matrix can be reliably estimated, even when the latter is very close to the identity matrix. We have illustrated our results on the example of stock returns correlations, that clearly reveal a non trivial structure for the bulk
eigenvalues. We have also discussed about the application to matrix denoising and we saw in particular that it is possible to make use of these overlaps between independent samples to approximate the so-called oracle estimator with great accuracy.

{\bf Acknowledgments:} We want to thank R. Allez, D. Bartz, R. B\'enichou, R. Chicheportiche, B. Collins, A. Rej, E. S\'eri\'e and D. Ullmo for very useful discussions.

\bibliography{overlap}

\newpage

\appendix 

\begin{widetext}

\begin{center}
	\b{\large{Appendix: Supplementary Material}}
\end{center}

\section{Mean squared overlap for independent sample covariance matrices}
\label{app:psi_SCM}

We keep the notations of Section \ref{sec:multiplicative_noise} and shall often omit the arguments $z$ and $\st z$ in $m$ and $\st m$ in the following when there is no confusion. Using the definition $m(z) = 1/(z\zeta(z))$, we rewrite \eqref{eq:psi_SCM} as
\begin{equation}
	\psi(z, \st z) \sim \frac{1}{1/\st m - 1/m} \left[ \frac{\stj}{\st z \st m} - \frac{\st \stj}{z m} \right],
\end{equation}
which is equivalent to
\begin{equation}
	\label{eq:m_v1}
	\psi(z, \st z) \sim \frac{1}{z\st z} \frac{1}{m - \st m} \left[ z \stj m - \st z \st \stj \st m \right].
\end{equation}
We can then express the function $\psi$ as a function of $m$ and $\st m$ only: 
\begin{eqnarray}
	\label{eq:m v2}
	\psi(z, \st z) 
	& \sim & \frac{1}{q \st q z \st z} \left[ \frac{(\st q z - q \st z) \st m^2}{m - \st m} + \frac{(q - \st q)\st m}{m - \st m} \right] + \frac{m+\st m}{q\st z} - \frac{1-q}{qz\st z}.
\end{eqnarray}
To use the inversion formula Eq.\ \eqref{eq:inversion_formula}, we need to evaluate $\psi(\lambda - \ii\eta, \st \lambda \pm \ii\eta)$. Using the short handed notation $m_0(\lambda) = \lim_{\eta\to0}m(\lambda - \ii\eta)$, we find 
\begin{multline}
	\label{eq:m_v3}
	\lim_{\eta\to 0} [ \psi(\lambda-\ii\eta, \st\lambda + \ii\eta) - \psi(\lambda-\ii\eta, \st\lambda - \ii\eta)] 
	\\ 
	\sim  \frac{\st q \lambda - q \st\lambda}{q\st q \lambda \st\lambda} \frac{m_0 \qB{\ovl{\st m_0^2} - \st m_0^2} + \st m_0 \ovl{\st m_0}\qB{{\st m_0} - \ovl{\st m_0}}}{\pB{m_0 - \ovl{\st m_0}} \pB{m_0 - \st m_0}} 
	+ \frac{\st q - q}{q\st q \lambda \st\lambda} \frac{m_0 \qB{\ovl{\st m_0} - \st m_0}}{\pB{m_0 - \ovl{\st m_0}} \pB{m_0 - \st m_0}} + \im,
\end{multline}
where we omitted the explicit expressions of the imaginary part since this is not important for Eq.\ \eqref{eq:inversion_formula}. 
Then, using the representation $m_0 = m_R + \ii m_I$ and $\st m_0 = \st m_R + \ii \st m_I$, one finds
\begin{eqnarray}
	m_0 \qB{\ovl{\st m_0^2} - \st m_0^2} + \st m_0 \ovl{\st m_0}\qB{\ovl{\st m_0} - {\st m_0}} 
	& = & 2 \st m_I \qb{ 2 m_I \st m_R + \ii (\st m_R^2 + \st m_R^2 - 2m_R\st m_R )}, \nonumber \\
	\ovl{m_0} \qB{\st m_0 - \ovl{\st m_0}}  & = & 2 \st m_I \qb{m_I - \ii m_R},
\end{eqnarray}
and
\begin{equation}
	\pB{m_0 - \ovl{\st m_0}} \pB{m_0 - \st m_0} = (m_R - \st m_R)^2 - (m_I^2 - \st m_I^2) + 2\ii m_I (m_R-\st m_R).
\end{equation}
Straightforward computations yields
\begin{eqnarray}
	\absB{ \pB{m_0 - \ovl{\st m_0}} \pB{m_0 - \st m_0} }^2 & = & \qB{ (m_R-\st m_R)^2 - (m_I^2 - \st m_I^2)}^2 + 4 m_I^2 (m_R-\st m_R)^2, \nonumber \\
	& = & \qB{(m_R - \st m_R)^2 + (m_I^2 + \st m_I^2)^2 }\qB{(m_R - \st m_R)^2 + (m_I^2 - \st m_I^2)^2},
\end{eqnarray}
which is exactly the denominator in \eqref{eq:mso_SCM_general}. For the numerator, elementary complex analysis in Eq.\ \eqref{eq:m_v3} yields
\begin{equation}
	\pB{m_0 \qB{\ovl{\st m_0^2} - \st m_0^2} + \st m_0 \ovl{\st m_0}\qB{\ovl{\st m_0} - {\st m_0}} } \times \ovl{ \pB{m_0 - \ovl{\st m_0}} \pB{m_0 - \st m_0} } = 4 m_I \st m_I \qB{ m_R \abs{\st m_0}^2 - \st m_R \abs{m_0}^2},
\end{equation}
and
\begin{equation}
	m_0 \qB{\ovl{\st m_0} - \st m_0 } \times \ovl{ \pB{m_0 - \ovl{\st m_0}} \pB{m_0 - \st m_0} } = 2 m_I \st m_I \qB{ \abs{\st m_0}^2  - \abs{m_0}^2}.
\end{equation}
By regrouping these last three equations with the prefactors in \eqref{eq:m_v3}, and recalling that $m_I(\lambda) = \pi q \varrho(\lambda)$ and $\st m_I(\lambdat) = \pi \st q \st \varrho(\lambdat)$, so we obtain by using the inversion formula \eqref{eq:inversion_formula} the following result:
\begin{equation}
	\Phi_{q,\st q}(\lambda, \lambdat) = \frac{2(\st q \lambda - q \lambdat) \qb{m_R \abs{\st m_0}^2 - \st m_R \abs{m_0}^2 } + (\st q - q) \qb{\abs{\st m_0}^2 - \abs{m_0}^2 }  }{\lambda\lambdat \qB{(m_R - \st m_R)^2 + (m_I + \st m_I)^2 }\qB{(m_R - \st m_R)^2 + (m_I - \st m_I)^2}}
\end{equation}
which is exactly Eq.\ \eqref{eq:mso_SCM_general}.

 \section{Mean squared overlap for deformed GOE matrices}
 \label{app:mso_dGOE}

The derivation of the overlaps \eqref{eq:mso_dGOE} for two independent deformed GOEs is very similar to sample covariance matrices. Hence, we shall omit most details that can be obtained by following the arguments of the above Appendix. Again, we shall skip the arguments $\lambda$ and $\st\lambda$ where there is no confusion. 

In Section \ref{sec:additive_noise}, we obtained 
\begin{eqnarray}
	\label{eq:psi_dGOE_v1}
	\lim_{\eta\to0}\qb{\psi(\lambda-\ii\eta,\st\lambda + \ii\eta) - \psi(\lambda-\ii\eta,\st\lambda - \ii\eta)} 
	& = & \frac{\stj_0\pB{\st \zeta^a - \ovl{\st \zeta^a}} + \zeta^a \pB{\ovl{\st \stj_0} - \st \stj_0} + \stj_0 \, \ovl{\st \zeta^a}  -  \ovl{\st \stj_0} \, \st \zeta^a }{\pB{\ovl{\st \zeta^a} - \zeta^a}\pB{{\st \zeta^a} - \zeta^a}}\,.
\end{eqnarray}
By proceeding as above (see Eq.\ \eqref{eq:m_v3} and thereafter), we find
\begin{equation}
	\stj_0\pB{\st \zeta^a - \ovl{\st \zeta^a}} + \zeta^a \pB{\ovl{\st \stj_0} - \st \stj_0} + \stj_0 \, \ovl{\st \zeta^a}  -  \ovl{\st \stj_0} \, \st \zeta^a = 2 \qb{ (\zeta^a_I \st \stj_I - \stj_I \st \zeta^a_I)  + \ii\pb{ \st\zeta^a_I (\stj_R - \st \stj_R) - \st \stj_I (\zeta^a_R - \st \zeta^a_R)}} 
\end{equation}
and
\begin{equation}
	\pB{\ovl{\st \zeta^a} - \zeta^a}\pB{{\st \zeta^a} - \zeta^a} = (\zeta^a_R - \st\zeta^a_R)^2  + \pB{\pb{\st \zeta^a_I}^2 - \pb{\zeta^a_I}^2} + 2 \ii \zeta^a_I \pB{\st \zeta^a_R - \zeta^a_R}.
\end{equation}
Hence, by putting these last two equations into \eqref{eq:psi_dGOE_v1} and then using \eqref{eq:inversion_formula}, we get after some straightforward computations
\begin{eqnarray}
	\Phi_a(\lambda, \lambdat) = \frac{ (\sigma^2 + \st \sigma^2)(\zeta^a_R - \st\zeta^a_R)^2 + 2\sigma^2\st \sigma^2 (\stj_R - \st \stj_R)(\zeta^a_R - \st \zeta^a_R) - (\sigma^2 - \st \sigma^2)((\zeta^a_I)^2 - (\st\zeta^a_I)^2) }{\qb{(\zeta^a_R -  \st \zeta^a_R)^2 + (\zeta^a_I +  \st \zeta^a_I)^2 } \qb{(\zeta^a_R -  \st \zeta^a_R)^2 + (\zeta^a_I -  \st \zeta^a_I)^2 } }\,,
\end{eqnarray}
which is exactly \eqref{eq:mso_dGOE} after some manipulations.

\section{The case of correlated Gaussian additive noises}
\label{app:mso_dGOE_corr}

In this section, we give the exact derivation of Eq.\ \eqref{eq:mso_dGOE} with correlated noises. Let 
\begin{equation}
	\label{eq:dGOE_correlated}
	\b S = \b C + \b W_1, \quad \St = \b C + \b W_2,
\end{equation}
where $\b W_1, \b W_2$ are two correlated GOE matrices (independent from $\b C$) satisfying
\begin{equation}
	\label{eq:correlation_structure}
	\avg{\b W_1}_{\cal N} = 0, \quad \avg{\b W_2}_{\cal N} = 0, 
	\quad \text{Cov}(\b W_1, \b W_2) = \begin{pmatrix}
		\sigma_1^2 & \rho_{12}
		\\
		\rho_{12} & \sigma_2^2
		\end{pmatrix}\,,
\end{equation}
where we denoted by $\cal N$ the Gaussian measure and used the abbreviation $\rho_{12} = \rho \sigma_1 \sigma_2$. Using the stability of GOE under addition, let us rewrite the noise terms as
\begin{eqnarray}
	\label{eq:noise_decomposition}
	\b W_1 & = & \b A + \b B_1, \nonumber\\
	\b W_2 & = & \b A + \b B_2, 
\end{eqnarray}
where $\b A$ that satisfies 
\begin{equation}
	\avg{\b A}_{\cal N} = 0, \quad \avg{\b A^2}_{\cal N} = \rho_{12}, 
\end{equation}
and $\b B_1$ and $\b B_2$ are two GOEs matrices independent from $\b A$ with
\begin{equation}
	\begin{cases}
	\avg{\b B_1}_{\cal N} = 0, \quad \avg{\b B_2}_{\cal N} = 0, \\
	\avg{\b B_1^2}_{\cal N} = \sigma_{1}^{2} - \rho_{12}, \quad \avg{\b B_2^2}_{\cal N} = \sigma_{2}^{2} - \rho_{12},\\
	\avg{\b B_1 \b B_2}_{\cal N} = 0.
	\end{cases}
\end{equation}
One can check that this parametrization yields exactly the correlation structure of Eq.\ \eqref{eq:correlation_structure}. Therefore, using \eqref{eq:noise_decomposition} into \eqref{eq:dGOE_correlated}, we have the equivalence (in law)
\begin{equation}
	\label{eq:dGOE_correlated_v2}
	\b S_1 = \b D + \b B_1, \quad \st{\b S} = \b D + \b B_2\,,
\end{equation}
where we defined 
\begin{equation}
	\b D \deq \b C + \b A \;\overset{\text{law}}{=} \;  \b C + \sqrt{\rho} \b W_0\,,
\end{equation}
with $\b W_0$ a GOE matrix with variance $\sigma_1 \sigma_2$. Since the noises are now independent and that the mean squared overlap $\Phi_a$, given in Eq.\ \eqref{eq:mso_dGOE}, is ``independent'' from the exact structure of $\b C$, we can therefore replace $\b C$ by $\b D$. Hence, we deduce that the overlaps for this model will again be given by Eq.\ \eqref{eq:mso_dGOE} with $\sigma^2 = \sigma_1^2 - \rho_{12}$, and $\st{\sigma}^2 = \sigma_2^2 - \rho_{12}$, as announced. \\

	

\end{widetext}

\end{document}